\documentstyle[12pt]{article}
\newcommand {\tr}{\mbox{ tr }}

\newcommand{\la}[1]{\label{#1}}
\newcommand{\be}{\begin{equation}}
\newcommand{\beq}{\begin{equation}\label}
\newcommand{\eeq}{\end{equation}}
\newcommand{\ee}{\end{equation}}
\newcommand {\bear}{\begin{eqnarray}}
\newcommand {\ear}{\end{eqnarray}}

\begin{document}
\setcounter{footnote}{0}
\renewcommand{\thefootnote}{\fnsymbol{footnote}}
\vspace{1cm}
\begin{center}
{\bf\Large
Chiral expansion in
the dual (string) model of the Goldstone meson
scattering\footnote{
Work supported by the Russian Foundation for
Basic Researches, grant 96-02-18017.} } \\
\vspace{1cm}
\setcounter{footnote}{0}
\renewcommand{\thefootnote}{\arabic{footnote}}
{\bf\large
M.V.~Polyakov{\footnote{e-mail:maxpol@thd.pnpi.spb.ru}} }\\
{\it Petersburg Nuclear Physics Institute, Gatchina,\\
St.Petersburg 188350, Russia} \\[.1cm]
{\bf\large
G.~Weidl{\footnote{e-mail:galia@vanosf.physto.se}}  }\\
{\it Institute of Theoretical Physics\\
 University of Stockholm, \,\, Box 6730\\
 S-113 85 Stockholm, Sweden}
\end{center}

\begin{abstract}
We consider in details the dual models for the Goldstone mesons
(pions) scattering in the presence of the explicit chiral symmetry
breaking caused by non--zero current quark mass. New method of
incorporation of the quark masses into the dual model is suggested.
In contrast to the previously considered in the literature methods, the
dual amplitude obtained by this method  is consistent with all
low--energy theorems following from the  Effective Chiral Lagrangian
(EChL) to
the $O(p^4)$ order and simultaneously it does not contain states with
negative width. The resonance spectrum of the model and its
implications for the fourth and sixth order EChL in large $N_c$ limit
are discussed. We argue that the possible relations between large $N_c$
QCD and some underlying string  theory can be revealed by studying
interactions of hadrons at low--energies.
\end{abstract}

\newpage
\noindent

\renewcommand{\thefootnote}{\arabic{footnote}}

\noindent
{\bf 1. }The success of the Regge phenomenology and applications of the
finite energy sum rules led in the late sixties to the formulation of the
duality principle for the hadronic amplitudes \cite{DHS} as a certain
relation between two ways of describing scattering amplitudes: the Regge
pole exchange at high energies and resonance dominance at low
energies.  Almost immediately it was found that the duality arises
naturally in the quantum theory of extended objects -- strings.
In other words the string theory appeared originally to describe some
striking features of strong interactions.
With the advent of the Quantum Chromodynamics it was
 found that QCD in the large colour number limit  $N_c\to \infty$ shares
many qualitative features with string theories. Let us just
(sporadically) list the evidences of deep relations between QCD and some
string theory :

\begin{itemize}
 \item The perturbation expansion in the large $N_c$
limit of  QCD can be written as a sum over surfaces
which may correspond to a sum over string world sheets
\cite{tHooft74}
\item
The strong coupling expansion for lattice gauge theory strongly resembles a
string theory
 \cite{lattice}
\item
The Wilson loop expectation values in the large $N_c$ limit
satisfy equations which are equivalent to those for one or
another specific string theory strings \cite{Migdal_Makeenko}
\item
2D QCD can be rewritten as a string theory \cite{Gross}
\end{itemize}
All these facts instill belief that QCD in the large $N_c$ limit
corresponds to some string theory, though the
particular form of such theory is not found.  This task is
especially
difficult because it involves comparing field theory (QCD) in
which we can not compute hadron amplitudes, with a string
theory in which basically all one can do is to compute
$S$-matrix in the narrow resonance approximation.
Probably the most natural place to make such a comparison is a low energy
region, where the dynamics is governed by soft pions. For the latter both
QCD and dual (string) models made  quantitative predictions.

Another portion of facts in favour
of possible underlying string dynamics
can be found in regularities of hadronic spectrum and
interactions at low energies. We again give only a short (incomplete)
list of them:

\begin{itemize}
 \item
The relative meson/baryon abundance might be explained by
string modular invariance \cite{Centr}.
\item
The Ademolo, Veneziano and Weinberg (AVW) mass relations
found in early days of string theory for
hadrons linked by the $S$--wave pion emission
$\alpha^\prime (M_\ast^2-M^2)=1/2$~\mbox{\rm mod(1)}
 \cite{AVW} was rederived from general structure and properties
of conformal field theory amplitudes \cite{Lew}.
\item
The Weinberg's mended symmetries
(algebraic realization of the chiral symmetry)
\cite{Weinber_alg_real,Weinberg_mended_symmetry} can be obtained as
a consequence of Kac--Moody algebra on the string world sheet
\cite{mylec}.
\item
The effective chiral lagrangian to the $O(p^6)$ order obtained by
the low--energy reduction of the dual resonance model for pions is in
agreement with phenomenology and non--topological chiral anomaly of QCD
\cite{MV}
\footnote{Recent general discussion of the
relation between chiral symmetry
and duality see in ref.~\cite{Vernew}}.

\end{itemize}

In this paper we study in details the dual 4--point amplitude for the
(pseudo)Goldstone mesons. The main our concern is a proper inclusion
of the current quark masses into the model, $i.e.$ effects related to
the deviations from the chiral limit. We require that the sought--for
amplitude  satisfies the following conditions:
\begin{itemize}
\begin{enumerate}
\item
In the chiral limit it is reduced to the Lovelace--Shapiro amplitude
\cite{Lovelace,Shapiro} and inclusion of the mass does not spoil
duality properties of the Lovelace--Shapiro amplitude.
\item
The low--energy limit of the amplitude is consistent with the effective
chiral lagrangian to the $O(p^4)$ order.
\item
There are no ghosts (states with negative residue).
\end{enumerate}
\end{itemize}

In the late sixties Lovelace \cite{Lovelace} and  Shapiro \cite{Shapiro}
proposed the following $\pi\pi$ scattering amplitude (in the chiral
limit)

\bear
\nonumber
M^{abcd}&=& \tr(\tau^a\tau^b \tau^c \tau^d) V(s,t) + \mbox{non-cyclic
permutations}, \\
V(s,t)&=& \lambda
\frac{\Gamma(1-\alpha_\rho(s))\Gamma(1
-\alpha_\rho(t))}{\Gamma(1-\alpha_\rho(s)-\alpha_\rho(t))},
\label{Lovelace_amplitude}
\ear
where the $\rho$-meson Regge trajectory and the constant $\lambda$ are
chosen to be
 \bear \nonumber
 \alpha_\rho(s)&=&\frac{1}{2} + \frac{s}{2 M_\rho}, \\
\lambda&=&-\frac{M_\rho^2}{\pi F_0^2},
\ear
in order to ensure the low-energy theorem for the amplitude:
\beq
\protect{\lim_{s,t \to 0}} V(s,t)=\frac{s+t}{F_0^2} + O(p^4),
\eeq
where $F_0 = \lim_{m_q\to 0} F_\pi$ is a pion decay constant in the
chiral limit.
The amplitude (\ref{Lovelace_amplitude}), besides correct low--energy
properties, satisfies a Regge asymptotic restrictions at high energies
and the positions and residues of the resonance poles are in a good
agreement with the phenomenological ones. Hence the basic phenomenological
features of the hadron interactions are implemented by a simple dual
amplitude (\ref{Lovelace_amplitude}). Let us stress that the amplitude
(\ref{Lovelace_amplitude}) could not serve for the $\pi\pi$ scattering
amplitude in the real QCD, if for no other reason than severe
violation of unitarity.
It could be rather considered as an
amplitude for some limiting case (large $N_c$) of the real QCD.
Indeed, the amplitude (\ref{Lovelace_amplitude}) possess all
properties peculiar to the hadronic amplitudes in the large $N_c$
limit: it has only pole singularities on the real axis of kinematical
variables, the OZI rule is absolute. In the present paper we shall
confront the dual models for pions not with the experimental data but
with the low--energy theorems  at large $N_c$ limit.

Recently the amplitude (\ref{Lovelace_amplitude}) and its many
point generalization was derived in the composite superconformal string
model suggested by V.~Kudryavtsev \cite{ku}. Although the chiral
properties of this model are still not studied.

In  pioneering works of Lovelace \cite{Lovelace} and
Shapiro \cite{Shapiro} the departure from the strict chiral limit was
achieved by shifting the intercept of the $\rho$-meson Regge trajectory
from $\frac{1}{2}$ to $\frac{1}{2}-\frac{M_{\pi}^2}{2M_{\rho}^2}$
to reproduce the Adler
zero of the off--shell amplitude.  Let us stress that the Adler
condition is an {\it off mass shell} one, whereas the dual (string)
amplitudes can be defined and constructed consistently only {\it on
 mass shell}, and a continuation of those to unphysical region is
 ambiguous.  Another problem is the appearance in this scheme
 of states
 with negative width.  Following ref.~\cite{MV} we shall use a new way
of introducing quark masses (explicit chiral symmetry breaking) into
 the dual model.  Instead of using the Adler condition  we shall impose
{\it on mass shell} low energy theorems on the dual amplitudes for the
$\pi\pi$ scatterings.

\vspace{0.3cm}
\noindent
{\bf 2. }
In the lowest order of momentum expansion $O(p^2)$ the interactions
of (pseudo)Goldstone mesons (pions, kaons and eta mesons) are
described by the famous Weinberg lagrangian \cite{WeiOld,Wei}:

\be {\cal L}^{(2)}=\frac{F_0^2}{4}
\tr(L_\mu L^\mu) + \frac{F_0^2 B_0}{4} \tr(\chi),
\la{echl2}
 \eeq
where $\chi=2 B_0 (\hat{m}U+U^\dagger\hat{m} )$, $L_\mu=iU\partial_\mu
U^\dagger$,
$\hat{m}=\mbox{diag}(m,m,m_s) $ is a quark mass matrix and
$F_0$ and $B_0$ are low-energy coupling constants carrying
an information about long-distance behaviour of the QCD.
The latter are related to the pion decay constant and the quark
condensate in the chiral limit as follows
\bear
\nonumber
F_0 &=& \lim_{m_q \rightarrow 0}F_\pi \approx 88\, \mbox{\rm MeV}\, , \\
\nonumber
B_0 &=& -\lim_{m_q \rightarrow 0} \frac{\langle \bar \psi \psi
\rangle}{F_\pi}.
\ear
The chiral field $U(x)$ is a unitary $3 \times 3$ matrix and
is parametrized in terms of eight pseudoscalar meson fields
$\pi$, $K$ and $\eta$:

\bear
\nonumber
U(x)&=& e^{i\Pi}, \\
\Pi&=&\left(\begin{array}{ccc}
  \frac{\pi^0}{F_\pi}+\frac{\eta}{\sqrt3 F_\eta} &-\sqrt2
 \frac{\pi^+}{F_\pi} & -\sqrt2 \frac{K^+}{F_K}\\
  -\sqrt2 \frac{\pi^-}{F_\pi}& -\frac{\pi^0}{F_\pi}+
  \frac{\eta}{\sqrt3 F_\eta} & -\sqrt2 \frac{K^0}{F_K}\\
  -\sqrt2 \frac{K^-}{F_K}&-\sqrt2 \frac{\bar K^0}{F_K} & -
  \frac{2\eta}{\sqrt3 F_\eta}
                 \end{array} \right),
\ear
with decay constants normalized as $F_\pi=93.3$~MeV,
$F_K\approx 1.2\,\, F_\pi$.

In the next $O(p^4)$ order the interactions of the (pseudo)Goldstone
mesons are described by the following EChL (we write only terms surviving
in the large $N_c$ limit and contributing to $\pi\pi$ scattering
amplitude) \cite{GLsu3}:

  \bear {\cal L}^{(4)}&=& (2L_2+L_3)\tr(L_\mu L^\mu
 L_\nu L^\nu ) + L_2  \tr(L_\mu L_\nu L^\mu L^\nu ) \nonumber \\ &+&
L_5 \tr(L_\mu  L^\mu \chi)+ L_8\tr(\chi^2).  \la{echl4} \ear
For the parameters of the  fourth order  EChL we use
here notations of Gasser and Leutwyler \cite{GLsu3}. We see that
the fourth order EChL has in the large $N_c$ limit four
independent parameters.\footnote{Without taking the large $N_c$
limit it depends on eight parameters \cite{GLsu3}.}

The last decade was a significant progress in understanding  the
structure of the fourth order EChL. In particular, the values of the
corresponding constants was determined from the experiment
\cite{GL,GLsu3,BijColGas} with a good accuracy,
these values were also
calculated in various approaches to the low--energy QCD
\cite{DiaEid,Bal,And,DiaPet,PraVal,Bij}. All these calculations relied
on large $N_c$ limit.

Let us consider the elastic $\pi \pi$--scattering process

$$
\pi_a(k_1)+\pi_b(k_2)\rightarrow\pi_c(k_3)+\pi_d(k_4)
$$
( $a,b,c,d=1,2,3$ are the isotopic indices and
$k_1,..,k_4$ --- pion momenta).
Its amplitude $ M^{abcd} $  can be written in the form:

\be
\protect{M^{abcd}}=\delta^{ab}\delta^{cd}A+\delta^{ac}\delta^{bd}B+
\delta^{ad}\delta^{bc}C ,
\label{A-def}
\eeq
where $A,B,C$ are the scalar functions of Mandelstam variables
 $s,t,u$:

\be
\protect{s}=(k_1+k_2)^2,\qquad  t=(k_1-k_3)^2,\qquad  u=(k_1-k_4)^2,
\la{mand} \eeq
obeying the Bose--symmetry requirements:

\begin{eqnarray}
A(s,t,u) &=& A(s,u,t) \ , \nonumber \\
B(s,t,u) &=& A(t,s,u) \ , \label{A-def2} \\
C(s,t,u) &=& A(u,t,s) \ . \nonumber
\end{eqnarray}
At low momenta one can expand the (iso)scalar
amplitude $A$
in power series of invariant kinemantical variables around
point $s=0, \, t=0$:

\be
\protect{A(s,t)}=\sum_{i,j}^{} \frac{A_{ij}}{F_0^{2 (i+j)}} s^i t^j .
\label{A-exp} \eeq
The expansion coefficients
$A_{ij}$ can be computed as a series in the quark mass $m$ with
help of EChL (\ref{echl2}), (\ref{echl4}). The tree level (large
$N_c$) computations give:

\bear
\nonumber
A_{00}&=&-\frac{2mB_0}{F_0^2}
+ \frac{64 m^2 B_0^2}{F_0^4}  (3L_2+L_3) +O(m^3)\, ,
\\
\label{a01:let}
A_{10}&=&1 -\frac{32 m B_0
}{F_0^2}(2L_2+L_3)+O(m^2) \, , \\
\nonumber
A_{20}&=&4(2L_2+L_3)+O(m)\, ,\\
\nonumber
A_{02}&=&8 L_2+O(m) \, .
\ear
Non--analytical contributions to these low--energy theorems appear due
to the loop contributions and hence suppressed by $1/N_c$.
We shall use the low--energy theorems at large $N_c$
eqs.~(\ref{a01:let}) to restrict the parameters of the dual models and
simultaneously to determine the parameters $L_i$ of the fourth order
EChL. Let us stress that the low--energy theorems (\ref{a01:let}),
in contrast to the Adler zero conditions, deal with the on--mass shell
amplitude.

\vspace{0.3cm}
\noindent
{\bf 3. }
In the presence of the small explicit chiral symmetry breaking (non-zero
quark mass $m$) we can generically write down the dual
amplitude for pions as a series in the powers of the quark mass $m$:

\bear
\nonumber
V(s,t) = -\frac{M_\rho^2}{2 \pi F_0^2}
(1+a_1 m +a_2 m^2 +\ldots)
\biggl\{
\frac{2\Gamma(1-\alpha_\rho(s))
\Gamma(1-\alpha_\rho(t))}{\Gamma(1-\alpha_\rho(s)-\alpha_\rho(t))}
\nonumber\\
+\sum_{(k,n,p)\neq (1,1,1)}(b_1^{(k,n,p)} m + b_2^{(k,n,p)} m^2+ \ldots)
[\frac{\Gamma(n-\alpha_\rho(s))
\Gamma(k-\alpha_\rho(t))}{\Gamma(p-\alpha_\rho(s)-\alpha_\rho(t))}+
(k \leftrightarrow n)]
\biggr\},
\label{pi}
\ear
where $k, n, p$ are natural numbers, such that

\beq{knp}
1 \leq n \leq k \leq p \leq n+k.
\eeq
The conditions (\ref{knp}) ensure that satellites we add
do not spoil the dual
properties of the resulting amplitude \cite{Shapiro}.
The indices $k,n,p$ label the unknown expansion
coefficients $b^{(k,n,p)}_i.$
The latter will be fixed by use of the low-energy theorems
(\ref{a01:let}) and by a  requirement of positivity of the
resonance widths.
The intercept of the $\rho$-meson Regge trajectory has also mass
corrections:

\beq{0}
\alpha_\rho(x)=\frac{1}{2}
(1+ i_1 m+ i_2 m^2+\ldots)+\frac{x}{2 M_\rho^2}.
\label{rhos_trajectory}
\eeq
$M_\rho$ is the mass of the $\rho$ meson in the chiral limit
and the coefficients $i_k$ enter with the mass corrections to the
intercept of the
$\rho$ meson trajectory and simultaneously the quark mass
corrections
{\footnote{We do not include corrections to the slope of
the trajectory because they can be absorbed by a redefinition of
the $\rho$-mass.}}
to the $M_\rho$.  Using the $SU(3)_{fl}$ symmetry it is
easy to relate the coefficient $i_1$ to the linear in strange quark
mass $m_s$ correction to the $K^*(892)$ mass (assuming $m_u=m_d=0$):

\be
M_{K^*}^2=M_\rho^2(1-\frac{m_s i_1}{2}+ O(m_s^2)).
\label{kastmass}
\eeq

In (\ref{pi})  all quantities are understood
as a chiral  expansion in the quark mass $m$.
In the chiral limit the amplitude
(\ref{pi}) coincides with
(\ref{Lovelace_amplitude}).

\vspace{0.3cm}
\noindent
{\bf 4. }
The partial width of the spin-$l$ resonance of mass $M_r(N)$ with
isospin $I$ is given by

\beq{rw}
\Gamma_I(N,l)=
-k_I \frac{q_{c.m.}}{16\pi^2 M_r^2(N)}
\int_{-1}^1 dz \,\, P_l(z) \,\,  \gamma(N,z),
\eeq
where
$$k_I=\biggl\{\begin{array}{ll} 3/2, & I=0\\ 1, & I=1\\ 0, &
I=2,\end{array}$$
$P_l(z)$ are the Legendre polynomials,
the $q_{c.m.}$ is the c.m. momentum  given by
\beq{qcm} q_{c.m.} =\frac{1}{2} \sqrt{M^2_r(N)-4 M_{\pi}^2},
\eeq
and the residue of the pole at $\alpha(s)=N$ is given by
\bear
\gamma(N,z)&=& {\mbox{Res}}_{\alpha(s)=N}V(s,t)=  \nonumber \\
&=&-\frac{ 2M_{\rho}^4}{\pi F_0^2}(-1)^{N+1}(1+a_1 m +a_2 m^2
+\ldots) \Bigl\{
\frac{-2\Gamma(1-\alpha_\rho(t))}{\Gamma(N)\Gamma(1-N-\alpha_\rho(t))}
\nonumber \\
&+&\sum_{(k,n,p)\neq (1,1,1)}(b_1^{(k,n,p)} m + b_2^{(k,n,p)} m^2+
\ldots) \\
&\times&  [\frac{(-1)^n \Gamma(k-\alpha_\rho(t))}{\Gamma(N-n+1) \Gamma(p-N-\alpha_\rho(t))}
+ (k
\leftrightarrow n)] \Bigr\},
 \label{g}
\ear
where the Mandelstam variables are expressed in terms of cms momentum
$q$ and cms scattering angle $z=\cos \theta$:
\beq{eq:kin}
   s=4(q^2+M_\pi^2),\qquad   t=-2q^2(1-z),\qquad  u=-2q^2(1+z).
\eeq
The mass of the resonances associated with this
pole is given by
\beq{mr}
M^2_r(N)={2 M_\rho^2}(N-\frac{1}{2}(1+ i_1 m+ i_2 m^2+\ldots)).
\eeq

The positivity of the resonance widths can be formulated as a
positivity of the following integral:

 \beq{ex} -\int_{-1}^1 dz \,\, P_l(z) \,\,
\gamma^{I}(N,z) = I^{(0)}(N,l)+m I^{(1)}(N,l)+ \ldots , \eeq
 where
$I^{(0)},\,\,I^{(1)}$ denote the expansion coefficients in the quark
mass. It is known that for any $N$ and $l$ all integrals $I^0(N,l)$ are
non-negative \cite{Shapiro}. In order to ensure the positivity of the
widths (absence of ghost) one should satisfy at least two conditions:
\begin{enumerate}
\item
$I^{(1)}(N,l) \geq 0$  for $N$ and $l$ such that $I^{(0)}(N,l)=0$
(if the $I^{(0)}(N,l)$ is positive then the chiral corrections
are not able to change a sign of the leading contribution
$I^{(0)}(N,l)$ ).  Actually $I^{(0)}(N,l)=0$ only for $N=2$ and
$l=0$ \cite{Shapiro}.
\item
The ratio  $I^{(1)}(N,l)/I^{(0)}(N,l)$ should not rise with
$N \to \infty $ at fixed $l$, otherwise one could not rely on
chiral counting for states with mass $\sim M_\rho^2/m$.
\end{enumerate}
We checked that the second condition is satisfied only for
$b_i^{112}, b_i^{212}$, and $ b_i^{213}$ satellites.
For the others the ratio $I^{(1)}(N,l)/I^{(0)}(N,l)$ grows with
$N$ and moreover oscillates in sign.

Furthermore the satellites $b_i^{112}, b_i^{212}$, and $ b_i^{213}$ are
not independent due to the identities

$$C^{112}=2 C^{213}$$
$$C^{111}+C^{112}-2 C^{212}=0$$
where
\beq{C}
C^{knp} \equiv [\frac{\Gamma(n-\alpha_\rho(s))
\Gamma(k-\alpha_\rho(t))}{\Gamma(p-\alpha_\rho(s)-\alpha_\rho(t))}+
(k \leftrightarrow n)].
\eeq
This observation leaves us with the following form of the generalized
dual amplitude

\bear
\nonumber
V(s,t)&=&-\frac{m_\rho^2}{\pi F_0^2}(1+a_1 m +a_2 m^2 +\ldots)
\biggl\{
\frac{\Gamma(1-\alpha_\rho(s))\Gamma(1
-\alpha_\rho(t))}{\Gamma(1-\alpha_\rho(s)-\alpha_\rho(t))} \\
&+&(b_1^{112} m + b_2^{112} m^2+ \ldots)
\frac{\Gamma(1-\alpha_\rho(s))\Gamma(1
-\alpha_\rho(t))}{\Gamma(2-\alpha_\rho(s)-\alpha_\rho(t))}
\biggr\}.
\label{pipi_parameters_free}
\ear
This amplitude was used in \cite{MV} in order to determine the
parameters of the fourth and sixth order EChL.
Here we address additionally the problem of the resonance widths, in
particular, the absence of ghosts.

It is known that the Lovelace-Shapiro amplitude does
not contain ghosts \cite{Shapiro}. Including the mass corrections
(\ref{0}) to the intercept or addition of satellites to
(\ref{Lovelace_amplitude}) may cause appearance of ghosts.
Since all chiral corrections are proportional to the quark mass, they
are responsible only for the small mass corrections and therefore they
can not compensate the contributions from the leading term
$\Gamma_I^{(0)}(N,l)$ when it is positive and non--zero.
Thus the problem of the positivity of the resonance widths is
equivalent to
the study of the mass corrections $I^{(1)}(N=2,l=0)$, the case when the
leading order result is zero.

Taking into account the chiral expansion at large $N_c$ of the pion mass
\beq{.}
M_{\pi}^2=2mB_0 +O(m^2),
\eeq
we obtain
from (\ref{rw}) and (\ref{pipi_parameters_free})
for the resonance width $\Gamma_{I}(N=2,l=0)$
to the linear order of the chiral expansion
$$\Gamma_{I}^{(0)}(N=2,l=0)=0\, ,$$
\bear
\Gamma_{I}^{(1)}(N=2,l=0)=\frac{ M_{\rho}^3}{4 \pi F_0^2}
({i_1}+b_1^{(112)}  ),
\label{wid}
\ear
and we require this quantity to be positive, {\it i.e.}
${i_1}+b_1^{(112)} \geq 0.$

The unknown constants in (\ref{wid}) are fixed by use of the low-energy
theorems (\ref{a01:let}).
Namely, at low momenta one can expand the amplitudes in power series
of invariant kinemantical variables.  Then we compare the low-energy
expansion coefficients $A_{ij}$ from the dual resonance amplitude
(\ref{pipi_parameters_free})

\bear
A_{00}&=&\frac{M_\rho^2 m}{ F_0^2}(i_1 - b_1^{112}) + O(m^2)\, , \\
A_{10}&=&1+\frac{m}{2 }\Bigl[2\,a_1+8\,\log (2)\,i_1
+\frac{16 \log (2) B_0}{M_{\rho}^2}
- 4\log (2)\, b_1^{112} \Bigr]+ O(m^2)\, , \\
A_{20}&=&0+O(m) \, ,\\
A_{02}&=&\frac{F_0^2}{M_\rho^2 } \log(2)+O(m) \, ,
\ear
with the low--energy theorems (\ref{a01:let}).
This comparison gives the following results for the parameters of the
fourth order EChL \cite{BolManPolVer,MV}

\bear L_2&=&2L_1, \label{nc} \\
L_3&=&-2 L_2
\label{anomaly} \\
L_2&=& \frac{F_0^2}{8 M_\rho^2}
\log(2)\approx 1.25\times 10^{-3}.
\label{l2value}
\ear
The relation (\ref{nc}) is identical to the one following
from the large $N_c$ conditions for the meson scattering
amplitude \cite{GLsu3}.  These conditions are ``built in'' in the
dual resonance models through the Chan--Paton isotopic factor.
The relation (\ref{anomaly}) is exactly the
relation predicted by integration of the non-topological chiral
anomaly \cite{DiaEid,And,Bal,DiaPet}. Actually the chiral
anomaly relation $2L_2+L_3=0$ holds in the dual model due to the
existence of the zero trajectories in these models. These trajectories
arise in the dual models to prevent a double pole in the amplitude.
The single-term amplitude (\ref{Lovelace_amplitude}) has
simultaneous $s$- and $t$-channel poles. Hence an intersection zero
must occur to prevent a double pole. This is provided by the lines of
zeros:
\be
\alpha_\rho(s)+\alpha_\rho(t)=l\geq 1.
\eeq
The zero trajectory with $l=1$ is associated with the Adler zero
(in the chiral limit, we are considering now, the Adler zero is  located in
the physical region), so that, in a sense, one has in the dual model
not only the Adler zero but rather a line of zeros, what influences
higher order EChL. These zero patterns have been studied by Odorico
\cite{Odo} who suggested that they may not rely on a
specific dual model, but have a more general nature.

Moreover the numerical value
of $L_2$ (\ref{l2value}) is close to that found by Gasser and Leutwyler
in ref.~\cite{GLsu3} $L_2=(1.7 \pm 0.7)\cdot 10^{-3}$, to recent
determination of this constant from analysis of the $K_{l4}$
decay \cite{BijColGas} $L_2=(1.35 \pm 0.3)\cdot 10^{-3} $ and
simultaneously to that obtained by integration of the
non--topological chiral anomaly $L_2=1.58 \cdot 10^{-3}$.
The values of the combination  $2L_2+L_3$ obtained
in refs.~\cite{GLsu3,BijColGas} are consistent with
zero\footnote{For the direct check of the relation $2L_2+L_3=0$
dictated by non-topological chiral anomaly of QCD and dual
(string) models one need to repeat the fitting procedure of
ref.~\cite{BijColGas} using, among others, variable $2L_2+L_3.$}.

Also the comparison with the low--energy theorems gives the
value of the parameters entering (\ref{pipi_parameters_free}),
{\it{i.e.}}

\beq{b112}
b_1^{112}=\frac{2B_0}{M_{\rho}^2} +i_1 \, ,
\eeq
\beq{a1}
a_1=-2\log\,(2)( \frac{2B_0}{M_{\rho}^2}+i_1) \, .
\eeq

Thus, inserting (\ref{b112}) into (\ref{wid})
 we obtain the following condition for the positivity of the
resonance width\
\beq{i}
i_1+\frac{B_0}{M_{\rho}^2} \geq 0.
\label{positivity}
\eeq
Let us stress that this condition differs from the one following from the
requirement of the Adler zero \cite{Lovelace}. In the latter case it is

\be
i_1+\frac{2B_0}{M_{\rho}^2} = 0,
\eeq
what evidently contradicts the condition (\ref{positivity}).

Using the relation of the $i_1$ to the $K^*(892)$ mass
eq.~(\ref{kastmass}) and the leading order expression for the
$K$--meson mass :
\be
M_K^2=m_s B_0 +O(m_s^2),
\eeq
one gets the following positivity condition:
\beq{poscond}
M_{K^*}^{(1)2} -M_{\rho}^2-M_K^2 /2 \leq 0.
\eeq
Here we introduce the notation:

\be
M_{K^*}^{(1)2} \equiv M_\rho^2(1-\frac{m_s i_1}{2}) \,.
\eeq
Using $M_\rho=760$~MeV and $M_K^2\approx m_s B_0 \approx
(495)^2$~MeV$^2$ we obtain  from (\ref{poscond}) the restriction $M_{K^*}^{(1)}
\leq 837$~MeV.  We see that the maximal possible mass $M_{K^*}^{(1)}$ is
less than the experimental value of the $K^*(892)$ mass (892~MeV).
However we should remember
that the value of $M_{K^*}^{(1)}$ differs from the $K^*(892)$ mass by
corrections of order $O(m_s^2)$ which can easily explain the difference
in $55$~MeV between $M_{K^*}^{(1)}=837$~MeV and $M_{K^*}=892$~MeV.

The widths of resonances can be now easily calculated.
For example,
the width of the $\rho$-meson in this model is given by
\beq{rho}
\Gamma(N=1, l=1)=\frac{ M_{\rho}^2}{8 \pi^2 F_0^2}
\frac{1}{3}q_{c.m.}
[1-m\{\frac{4B_0}{M_{\rho}^2}(\ln2 +2)+(2\ln2 +1)i_1\}],
\eeq
where $q_{c.m.} =\frac{1}{2} \sqrt{M^2_{\rho}-4 M_{\pi}^2}.$
The width of the $f_2$-meson is obtained as

\beq{f}
\Gamma(N=2, l=2)=\frac{ M_{\rho}^4}{8\pi^2 F_0^2}
\frac{3q_{c.m.}}{10 M_{f_2}^2}
[1-m\{\frac{4B_0}{M_{\rho}^2}(\ln2 +\frac{4}{3})+2
(\ln2 +\frac{1}{3})i_1\}],
\eeq
where $q_{c.m.} =\frac{1}{2} \sqrt{M^2_{f_2}-4 M_{\pi}^2}.$

The numerical values of masses and
widths for some other resonance states are shown in Table~1.
For numerical calculations we choose $M_\rho=760$~MeV, $F_0=88$~MeV
and $i_1=-\frac{B_0}{M_\rho^2}$, the latter value corresponds
to the maximal possible value of $M_{K^*}^{(1)}=837$~MeV.

\vspace{0.3cm}
\noindent
{\bf 5. }
Summarizing, we made a detailed study of the implications of spontaneous chiral
symmetry breaking for dual (string) models of Goldstone boson scattering.
We suggested a new method of inclusion of the explicit chiral symmetry
breaking by the non--zero current quark mass into the dual (string) model.
This method, in contrast to the one suggested in the original papers
\cite{Lovelace,Shapiro}, does not lead to appearance of resonances with
negative width (ghosts). Constructed by this method $4\pi$ dual amplitude
on the one hand respects all low--energy theorems for soft pions, on the other
hand it gives particular predictions for the higher orders of the
effective chiral lagrangian (EChL). The most, in our view, remarkable
prediction is the relations $2L_2+L_3=0$ for the parameters of the fourth
order EChL. This relation is \underline{exactly} the same as that followed
from the integration of the chiral non--topological anomaly
\cite{DiaEid,Bal,And} and instanton model of the QCD vacuum \cite{DiaPet}.
The deeper understanding of this amazing coincidence could shed a light on
possible relations between quark--gluon dynamics and strings.

Let us once more attract reader's attention to the idea that the natural
region to look for evidences  of `` QCD strings" (possible string theory
underlying QCD) is a soft pion physics. Both strings and QCD make
particular quantitative  predictions in this region and hence can be
confronted there. In other words one could search for `` string structure"
of chiral expansion.

To realize this program the efforts from three sides are needed. First,
more precise measurements of Goldstone boson scattering could give an
additional information on  the ``string" chiral counting
\footnote{The string effects are shadowed by loop $1/N_c$ corrections, but
the latter are under quantitative theoretical control of chiral perturbation
theory \cite{two-loops}.}. Also the more precise and detailed meson
spectroscopy ({\it e.g.} search for scalar state with mass $\sim 1300$~MeV
and with coupling constant to $\pi\pi$ vanishing faster then
$1/\sqrt{N_c}$ in the large colour number limit\footnote{Note the
remarkable resemblance to elusive glueball  !}) would help in quest
 of ``QCD strings".  Second, from the QCD side some
additional efforts in understanding of the sixth order EChL are
required in order to confront sensibly ``QCD chiral expansion" with the
``string" one.  The third, construction of the many pion string
amplitudes and study of their low--energy behaviour are also needed.
Actually the failure of attempts to generalize the Lovelace--Shapiro
amplitude to the many pions led to loosing the interest to the strings
as a theory of strong interactions.  Very promising achievements in
construction of such generalization were recently made by
V.A.~Kudryavtsev \cite{ku}, although the chiral properties of suggested
n--point pion amplitudes were not yet studied.

\section{Acknowledgements}
This work has been supported by the Russian Foundation for
Basic Researches, grant 96-02-18017.
We are grateful to V.~Kudryavtsev and V.~Vereshagin for fruitful
discussions of duality.
M.V.P. acknowledges the hospitality of the
Institute for Theoretical Physics~II at Bochum University and Cracow
Nuclear Physics Institute where part of the work has been done.

\newpage

\begin{table}
\begin{tabular}{|c|c|c|c|c|c|c|c|c|c|c|}
\hline
$J^P$   &770&1327 &1710 &2022 &2293 &2534 &2755 &2959 &3150 &3330 \\
\hline
$10^+$  & & & & & & & & & &                 2 \\
$9^- $  & & & & & & & & &                2 &1.5 \\
$8^+ $  & & & & & & & &               5 &4 &5.5  \\
$7^- $  & & & & & & &              5 &4 &5 &4.5  \\
$6^+ $  & & & & & &            11 &9 &10&10&9  \\
$5^- $  &    &    &   &   &12 &10 &9 &9 &7 &7.5  \\
$4^+ $  &    &    &   &29 &24 &18 &19&13&14&10  \\
$3^- $  &    &    &32 &27 &15 &17 &10&12&7 &9  \\
$2^+ $  &    &81  &73 &25 &35 &16 &22&12&16&10  \\
$1^- $  &93  &100 &15 &32 &11 &18 &8 &13&7 &10  \\
$ 0^+$  &493 &0   &68 &15 &34 &13 &23&11&17&9  \\
\hline
\end{tabular}
\caption{
Masses and $\pi\pi$ widths (in MeV) of resonances predicted by dual
model with mass corrections.  }
\end{table}

\end{document}